\documentclass[9pt,conference]{IEEEtran}
\IEEEoverridecommandlockouts
\usepackage{amsmath}
\usepackage{graphicx}
\usepackage{amssymb}
\usepackage{xcolor}
\usepackage{verbatim}
\usepackage{booktabs}
\usepackage{multirow}
\usepackage{threeparttable}

\def\BibTeX{{\rm B\kern-.05em{\sc i\kern-.025em b}\kern-.08em
    T\kern-.1667em\lower.7ex\hbox{E}\kern-.125emX}}

\makeatletter
 \let\old@ps@headings\ps@headings
 \let\old@ps@IEEEtitlepagestyle\ps@IEEEtitlepagestyle
 \def\confheader#1{%
 \def\ps@headings{%
 \old@ps@headings%
 \def\@oddhead{\strut\hfill#1\hfill\strut}%
 \def\@evenhead{\strut\hfill#1\hfill\strut}%
 }%
 \def\ps@IEEEtitlepagestyle{%
 \old@ps@IEEEtitlepagestyle%
 \def\@oddhead{\strut\hfill#1\hfill\strut}%
 \def\@evenhead{\strut\hfill#1\hfill\strut}%
 }%
 \ps@headings%
 }
 \makeatother

\confheader{%
 This paper has been accepted and will be presented at EUSIPCO 2019
 }

\begin{document}
\title{CNN-based Multichannel End-to-End Speech Recognition for Everyday Home Environments* \thanks{The work has been supported by MEXT Grant-in-Aid for Scientific Research (A), No. 15H01710, except for the contribution of Mitsubishi Electric Research Laboratories (MERL).}}
\author{\IEEEauthorblockN{Nelson Yalta$^1$, Shinji Watanabe$^2$, Takaaki Hori$^3$, Kazuhiro Nakadai$^4$, Tetsuya Ogata$^1$}
\IEEEauthorblockA{\textit{$^1$Waseda University, $^2$Johns Hopkins University, $^3$Mitsubishi Electric Research Laboratories, } \\
\textit{$^4$Honda Research Institute Japan}\\
nelson.yalta@ruri.waseda.jp}}

\maketitle
%

%

%
\begin{abstract}
Casual conversations involving multiple speakers and noises from surrounding devices are common in everyday environments, which degrades the performances of automatic speech recognition systems. 
These challenging characteristics of environments are the target of the CHiME-5 challenge. By employing a convolutional neural network (CNN)-based multichannel end-to-end speech recognition system, this study attempts to overcome the presents difficulties in everyday environments. 
The system comprises of an attention-based encoder--decoder neural network that directly generates a text as an output from a sound input.
The multichannel CNN encoder, which uses residual connections and batch renormalization, is trained with augmented data, including white noise injection. 
The experimental results show that the word error rate is reduced by 8.5\% and 0.6\% absolute from a single channel end-to-end and the best baseline (LF-MMI TDNN) on the CHiME-5 corpus, respectively. 

\end{abstract}
\begin{IEEEkeywords}
End-to-end speech recognition, Multichannel, Residual networks
\end{IEEEkeywords}
\section{Introduction}
\label{sec:intro}


Automatic speech recognition (ASR) enables the machines to understand human languages and follow human voice commands. Currently, the ASR system implemented with deep learning techniques improves its performance in near/far fields \cite{Hinton2012, Liu2014} for diverse environmental conditions \cite{Delcroix2014LinearPD}.  
Recently, an ASR system implemented with end-to-end models (see e.g., \cite{Chorowski:2014, Amodei:2015, Watanabe2017, chiu2018state}) has gained attention because unlike conventional ASR system, end-to-end models learn to directly map character sequences from acoustic feature sequences without any intermediate modeling, such as the acoustic model, pronunciation lexicon, and language models based on deep learning \cite{Hinton2012,Bo2017}.

The two major approaches of end-to-end models, particularly connectionist temporal classification (CTC) \cite{Graves:2006,Miao:2015} and attention-based models \cite{Chorowski:2014,Lu:2016} have achieved promising recognition results.  
CTC-based models \cite{Graves:2006} solve sequential learning problems based on Markov assumptions \cite{Miao:2015}. 
Whereas, attention-based models align between acoustic frames and decoded symbols by using an attention mechanism \cite{Chorowski:2014,Lu:2016}. 
Recent studies on end-to-end models have shown that compared to the individual performance of each approach, a joint CTC--attention model improves the recognition performance \cite{Watanabe2017, Dalmia}. 
The joint model trains an attention-based encoder with an attached CTC objective for regularization. 
Furthermore, the CTC objective is employed during the decoding phase to improve the model results \cite{Hori2017}. 

Although end-to-end models are comparable or even more advantageous than the conventional ASR systems \cite{Watanabe2017, chiu2018state}, it is nevertheless challenging to robustly recognize speech signals in noisy environments and with low resources (i.e., CHiME-5 task \cite{Barker}). The CHiME-5 task comprises the difficulties of casual conversion with overlapped sentences or unfinished utterances, noises from home appliances at a signal-to-noise ratio (SNR) between 5 and 20 dB, distant microphone speech, and a small training dataset of 40 h (i.e., low resources).
Most competitive systems, except for \cite{Dalmia}, in the fifth CHiME challenge employ conventional ASR methods with multichannel speech enhancement techniques \cite{Du:2018, Kanda:2018,Medennikov:2018,Doddipatla:2018}. %

This study addresses the challenging characteristics of the CHiME-5 challenge using an end-to-end ASR model. The challenge considers distant multi-microphone speech captured by four binaural microphone pairs and six Kinect microphone arrays and features two tracks, namely the single-array track and the multiple-array track. 
Herein, under the conditions mentioned earlier, we propose an extension of a joint CTC--attention model that uses residual connections for the CNN and accepts multichannel inputs to boost the speech recognition performance. In particular, our multichannel end-to-end approach focuses on a single-array track.

First, we explore the use of multichannel inputs \cite{Sainath2017,Ochiai17} for noisy environments under the fifth CHiME challenge scenario \cite{Barker} to train our model. Then, we boost the performance adapting the model to accept inputs with a different number of channels (binaural microphone and single array track), namely the parallel encoder. By doing this, the model has a larger training set with almost clean sound data provided by the binaural microphone that enriches possible input feature combinations. Finally, we evaluate several configurations for a joint CTC--attention model with an end-to-end toolkit called ESPnet \cite{Watanabe2018}.

This study presents extensions of a joint CTC--attention model. The performance was evaluated and compared to that of a conventional joint CTC--attention model. 
The introduced extensions are as follows:
\begin{itemize}
\item Parallel CNN encoder with residual connections \cite{Zhang:2017}. We employed the data from both microphones (i.e., Kinect and binaural) to improve the performance for noisy speech recognition. 
Furthermore, we observed that augmenting the data on the binaural side with white noise reduced the absolute word error rate (WER) by 4\%, and obtained better performance than employing dropouts in the CNN encoder. 
\item Batch renormalization \cite{Ioffe2017}. This normalization improves the training process for small mini-batches using the moving averages of the mean and the variance during training and inference. 
\item Multilevel language modeling (LM) \cite{Hori2017LM}. This modeling technique integrates the ability to model an open vocabulary ASR of a character-based LM with the strength to model large sequences of word-based LM. 
\end{itemize}
For the CHiME-5 corpus, the absolute WER of the proposed extensions for joint CTC--attention model  improved by 14\% compared to that of a standard joint model. The extensions are additionally evaluated in the AMI corpus \cite{Hain2007}.


\section{End-to-End ASR Overview}
\label{sec:overview}
In this section, we give an overview of end-to-end ASR. The framework employs a joint CTC--attention model that processes the audio features and generates text as an output. 
\subsection{Joint CTC--Attention Model}
\label{sbsec:jointatt}
The key idea of a joint CTC--attention model is to overcome 1) the conditional independence of the targets assumed in the CTC model and 2) the misalignments in the attention model caused by the noise in real-environment speech recognition tasks \cite{KimHW16}. A joint CTC--attention model uses a shared encoder to train an attention model encoder with a CTC objective function as an auxiliary task. This model uses the multi-task learning (MTL) framework to achieve the desired training.  \par 
For an audio input $X$ of length $N$, CTC will generate and output a sequence of shorter length $C =\{c_l \in \mathcal{S} |l=1,.,L\}$ for the $L$-length letter sequence with $L \leq N$ and a set of distinct characters $\mathcal{S}$. 
CTC generates an intermediate "blank" symbol, which represents the omission of the output label. 
This special symbol is introduced to generate a frame-wise letter sequence $Z = \{z_t \in \mathcal{S} \cup \text{blank}|t=1, ...,T\}$. 
Assuming conditional independence between each output, CTC models the probability distributions over all possible label sequences to maximize $p(C|X)$ as follows: 
\begin{equation}
p_{\text{ctc}}(C|X) \triangleq p(C|X) \approx \sum_{Z}\prod_{t}p(z_t|z_{t-1},C)p(z_t|X)p(C)
\label{eq:pctc},
\end{equation}
where $p(z_t|z_{t-1},C)$ and $p(C)$ are the label prior distributions; $p(z_t|X)$ represents the frame-wise posterior distribution and is modeled  using a deep encoder \cite{Hori2017}.

In contrast, an attention-based model does not assume any conditional independence assumptions for $p(C|X)$. 
The posterior probability $p(C|X)$ is directly estimated based on the chain rule:
\begin{equation}
p_{\text{att}}(C|X) \triangleq p(C|X) \approx \prod_{l}p(c_l|c_1, ..., c_{l-1}, X),
\label{eq:patt}
\end{equation}
where $p(c_l|c_1, ..., c_{l-1}, X)$ is represented as:
\begin{equation}
p(c_l|c_1, ..., c_{l-1}, X) = \text{Decoder}(r_l, q_{l-1}, c_{l-1}),
\label{eq:att1}
\end{equation}
\begin{equation}
r_l=\sum_t{a_{lt}h_t},
\label{eq:att2}
\end{equation}
where Decoder($\cdot$) $\triangleq$ Softmax(Lin(LSTM($\cdot$))) is a recurrent neural network (RNN) with a hidden vector $q_{l-1}$, a previous output $c_{l-1}$, and a letter-wise hidden vector $r_l$; $a_{lt}$ is the attention weight and represents a soft alignment obtained by a content-based attention mechanism with convolutional features \cite{Chorowski15}.

\label{sbsec:jointdec}
\begin{figure}
  \centering
  \includegraphics[width=85mm]{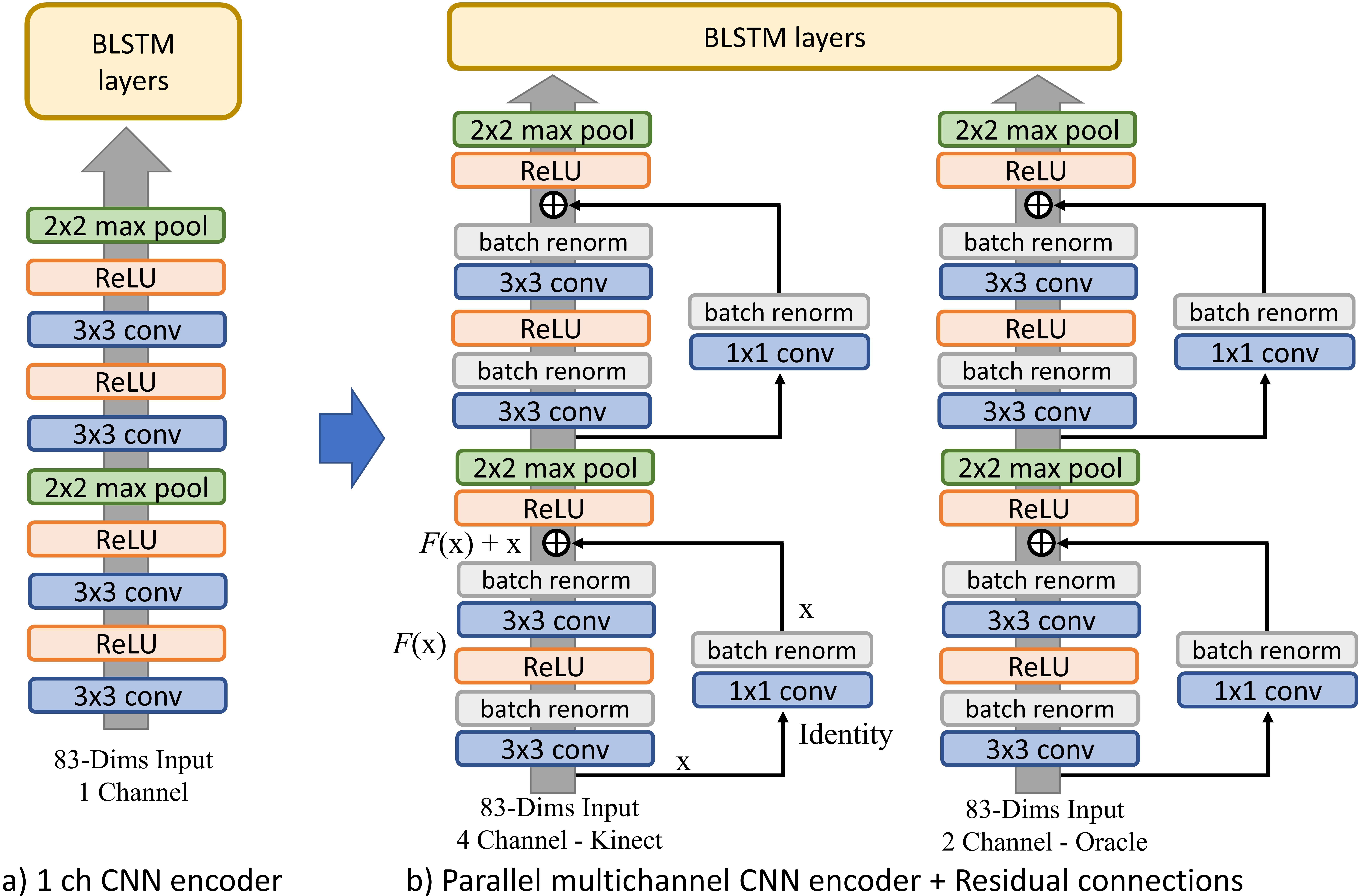}
  \caption{Parallel Encoder: From a joint CTC/attention model implemented with a) 1 channel (ch) CNN encoder, this is replaced by b) the parallel encoder which accepts inputs with a different number of channels.}
  \label{f:parallel}
\end{figure}

The use of a joint CTC--attention model with MTL approach improves the performance in the ASR task and reduces irregular alignments during training and inference. This MTL objective maximizes the logarithmic linear combination of the CTC and attention objectives:
\begin{equation}
\mathcal{L}_{\text{MTL}} =  \lambda \log p_{\text{ctc}}(C|X) + (1-\lambda) \log p_{\text{att}}(C|X),
\label{eq:mtl}
\end{equation}
where $\lambda$ is a tunable parameter with values $\lambda:0\leq \lambda \leq 1$.\par

\section{Adaptation for Multichannel ASR in Noisy environments}
\label{sec:adaptation}
The idea of our model is to use a parallel deep CNN encoder with residual connections, batch renormalization, and a multilevel RNN-LM network as an extension for a joint CTC--attention end-to-end ASR with multichannel input. The following subsections describe each individual extension in detail.

\subsection{Parallel Multichannel Encoder}
To boost the accuracy of the joint CTC--attention model applied in the fifth CHiME challenge, we employed both Kinect and binaural microphone arrays supplied on the corpus during training using a parallel multichannel encoder (Fig.~\ref{f:parallel}).
The multichannel encoder comprises of two CNNs that process each array during a mini-batch step and uses the CNN encoder with Kinect array during decoding because the binaural array cannot be used for the distant ASR scenario.
Unlike sole training with a single channel or multichannel from the Kinect array, using the binaural array enriches the possible input feature combinations and regularizes the network training, thereby improving the model performance.

\subsection{Residual Connections}
Using residual (i.e., skip) connections presents several benefits. 
They improve the back-propagation of the gradient to the bottom layers, thus easing the training on very deep networks \cite{He2015}. 
In a neural network, studies have shown that residual or skip connections eliminate the overlaps, consistent deactivation, and linear dependence singularities of nodes \cite{Orhan2017}. \par 
Let $H(x)$ be the learned mapping of a network. The network can then also learn $H(x)-x$ mapping for a given input $x$. Residual learning is then denoted as follows: 

\begin{equation}
H(x) := F(x) + x.
\end{equation}
Residual learning is implemented in any feedforward neural network using a skip connection (Fig.~\ref{f:parallel}), which is presented as an identity mapping. A network can be trained end-to-end with this implementation using any deep learning framework. In practice, this implementation improves model performance; thus, it increases the computing time.\par
In this study, residual learning is implemented using three convolutional layers, namely two convolutional layers with a kernel filter size of $3\times3$ for calculating $F(x)$ and one with a kernel filter size of $1\times1$, which is used as the skip connection. \par

\subsection{Batch Renormalization}
A recent technique, called batch normalization (BatchNorm) \cite{Ioffe2015}, has become the standard for the normalization process. BatchNorm computes the mean and variance of a mini-batch; furthermore, it normalizes the mini-batch with the computed values. In addition, the mean and variance are computed over all the training data to employ them for inference. However, the use of the mean and variance has a significant drawback when mini-batches with few samples are employed \cite{Ioffe2017}. \par
Batch renormalization \cite{Ioffe2017} proposes the application of a per-dimension affine transformation to the normalized activations. The statistic differences of a mini-batch are corrected by fixed parameters ensuring that the computed activations depend only on a single example; thus, the performance for models trained with small mini-batches is improved. Batch renormalization also employs the overall calculated mean and variance in the training process. During training, unlike batch normalization that uses the overall mean and variance only for inference, the above layers observe the same activations that would be generated for inference. \par
We boosted the accuracy of the joint model by implementing the model with batch renormalization in the CNN layers (Fig.~\ref{f:parallel}). 
This implementation improved the performance of the proposed models, thus obtaining an additional absolute error rate reduction of 0.1\% in the single-array track WER (Table~\ref{t:wer2}).  

\subsection{Multilevel RNN-LM}
Prior studies have shown that integrating the joint CTC--attention model with a character-based RNN-LM improves recognition accuracy \cite{Hori2017}. 
Word-based LM suffers from the out-of-vocabulary (OOV) problem, unlike the character-based LM that has the advantage of open vocabulary ASR \cite{Hori2017LM}. 
However, for the character-based LM, modeling linguistic constraints across a long sequence of characters is difficult. 
Previously, this problem was solved by implementing a multilevel LM and combining it with the decoder network \cite{Hori2017LM}.
Fist, the multilevel LM ranks the hypothesis using the character-based LM. Then, the word-based LM rescores known words. 
The OOV score is provided by the character-based LM.

\section{Experimental Setup}
\label{sec:setup}
We studied the effectiveness of our proposed extensions using the ESPnet speech recognition toolkit, which is an end-to-end speech processing toolkit \cite{Watanabe2018}, with Chainer backend \cite{Tokui2015}.  We present experiments with models training on 40 h of CHiME-5 data \cite{Barker} and 78 hours of AMI data  \cite{Hain2007}.  \par
The fifth CHiME challenge (CHiME-5) comprised tasks of conversational ASR employing distant multi-microphones in real home environments \cite{Barker}. The speech material captured natural and conversational speeches. Six Kinect microphone arrays and four binaural microphone pairs were employed to record it. The speech material comprised a total of 40 h of training data, 4 h of development data, and 5 h of evaluation data. The corpus features two challenges, namely the single-array track and the multiple-array track. Herein, we considered the single-array track (i.e., SAT). \par
The AMI dataset comprises tasks of speech recognition in meetings \cite{Hain2007}. The speech material was captured with 8-channel circular microphones (i.e., multiple distant microphone (MDM)), and a headset microphones (i.e., independent headset microphone (IHM)) and comprised approximately 78 h of training data and approximately 9 h of development and evaluation data. \par

Unless otherwise indicated, the experiments were performed using the parameters described in Table~\ref{t:config}.

We tested several values combinations of $\lambda$ for both training and decoding, where the values that are showed in Table~\ref{t:config} obtained lower WER.

\begin{table}[t]
\caption{Experimental configuration}
\label{t:config}
\centering

\begin{tabular}{ll}
\toprule[0.3mm]
\textbf{Feature} & \\
\begin{tabular}{@{}l@{}}Input stream\\(per channel)\end{tabular} & 80-dim fbank + 3-dim pitch \\
\hline
\textbf{Model} & \\
CNN-encoder type & VGG, Residual, Res+Batch Renorm.  \\
CNN-encoder layers & \begin{tabular}{@{}l@{}}VGG:4, Residual:6,\\  Res+Batch Renorm: 6\end{tabular}  \\
RNN-encoder type & BLSMTP \\
RNN-encoder units & 512 cells \\
RNN-encoder layers & 4  \\
Attention & Location-based \cite{Chorowski15} \\
Decoder type & 1-layer 300 cells LSTM \\
CTC weight $\lambda$ (train) & CHiME-5:0.1, AMI:0.5 \\
CTC weight $\lambda$ (decode)  & CHiME-5:0.1, AMI:0.3 \\
Optimization & AdaDelta \cite{Zeiler2012} \\
Epochs & 15 \\
\hline
\textbf{Character-based RNN-LM} & \\
Type & 2-layers 650 cells LSTM \\
Optimization & ADAM \cite{KingmaB14}  \\
\textbf{Word-based RNN-LM} & \\
Type & 1-layers 650 cells LSTM \\
Optimization & Adadelta  \\
\bottomrule[0.3mm]
\end{tabular}
\end{table}
\section{Experiments}
\label{sec:experiments}

We try to investigate the performance of each extension in the following subsections. 
In these experiments, we only report the WER(\%) results on the development set of CHiME-5 and on the development and evaluation sets of AMI. However, from Sections~\ref{sbsec:perturbation}, we only report the result for CHiME-5. 

\subsection{Single Channel Input}
As a preliminary experiment, we explored the ASR performance of a CNN-based encoder for the single-channel input. This experiment allowed us to adjust the training parameters for the experiments that follow. Table~\ref{t:wersinglech} presents the resulting WER. 

A subset of 275K utterances was randomly selected from both Kinect and binaural arrays to train a single-channel input model with CHiME-5. The single channel model employs a joint CTC--attention with a VGG-BLSTMP encoder. Unless otherwise stated, we use a character-based RNN-LM for decoding in subsequent sections. The result obtained was then compared to that reported in \cite{Barker}. The end-to-end baseline is a joint CTC--attention model implemented with a BLSMTP encoder and trained for 12 h. 

For AMI, the model was trained with each microphone array (i.e., IHM and MDM) separately. A single channel was synthesized using delay-and-sum beamforming \cite{Anguera:2007} to train the model with the MDM array (i.e., AMI-MDM). Unless otherwise indicated, a word-based RNN-LM is employed at the decoding stage in the consequent sections.  Furthermore, the results were compared to those found in the official webpage of
ESPnet\footnote{https://github.com/espnet/espnet/blob/master/egs/ami/asr1/RESULTS}. 
 
 \begin{table}
\caption{WER (\%) comparison for systems trained with a single channel input}
\label{t:wersinglech}
\centering
\begin{tabular}{lcccccc}
 & & \begin{tabular}{@{}c@{}}GMM\\\cite{Barker}\end{tabular} & 
 \begin{tabular}{@{}c@{}@{}}LF-MMI\\TDNN \\\cite{Barker}\end{tabular} &
\begin{tabular}{@{}c@{}}CMU\\\cite{Dalmia}\end{tabular} &
 \begin{tabular}{@{}c@{}@{}}End\\to\\End*\end{tabular} & \begin{tabular}{@{}c@{}@{}}CNN\\based\\Encoder\end{tabular}
 \\\cmidrule{3-7}
\multirow{2}{*}{CHiME-5} & SAT 
                            & 91.7 & \textbf{81.3} & 82.1 & 94.7 & 89.2 \\
                        & Binaural & 72.8 & \textbf{47.9} & - & 67.2 & 61.1 \\\hline
\multirow{2}{*}{AMI-IHM}    & dev  & - & - & - & 37.5 & \textbf{30.9} \\
                            & eval & - & - & - & 38.5 & \textbf{32.8} \\\hline
\multirow{2}{*}{AMI-MDM}    & dev  & - & - & - &  -   & \textbf{50.6} \\
                            & eval & - & - & - &  -   & \textbf{54.8} \\\hline
\end{tabular}
\begin{tablenotes}\footnotesize
\item[*]*Baseline \cite{Barker}
\end{tablenotes}
\end{table}

\subsection{Parallel Encoder}

In the first set of experiments, we explored the performance of the proposed multichannel CNN-based parallel encoder, particularly the parallel encoder. In Table~\ref{t:wermultich}, the WER for a single multichannel encoder (i.e., single encoder) and the parallel encoder are listed. With the parallel encoder, we can see a decrease in the WER on both datasets compared to that in the baseline single channel and the CNN-based encoder with a single-channel input. 

For CHiME-5, the single encoder employed four channels available on the single-array track. The parallel encoder had an input configuration of $4 + 2$. Four channels were available at the single-array track, and two channels were from binaural.

For AMI, the single encoder employed eight channels available on AMI-MDM. The parallel encoder had an input configuration of $8 + 1$. Eight channels were available on AMI-MDM, and one channel was from AMI-IHM.

\begin{table}
\caption{WER (\%) comparison for systems trained with multichannel input.}
\label{t:wermultich}
\centering

\begin{tabular}{lccc}
 & & \begin{tabular}{@{}c@{}}Single\\Encoder\end{tabular} &  \begin{tabular}{@{}c@{}}Parallel\\Encoder\end{tabular} \\
 \cmidrule{3-4}
\multirow{2}{*}{CHiME-5} &  SAT &  88.3 & \textbf{85.4} \\
& Binaural & - & \textbf{55.6} \\\hline
\multirow{2}{*}{AMI-IHM}    & dev & - & \textbf{29.4}  \\
                            & eval &　- & \textbf{30.1} \\\hline
\multirow{2}{*}{AMI-MDM}    & dev & 50.6 & \textbf{45.3} \\
                            & eval & 54.9 & \textbf{49.0}  \\\hline
\end{tabular}
\end{table}
\begin{table}
\caption{WER (\%) comparison for CNN-based architectures of the Parallel Encoder.}
\label{t:wer2}
\centering

\begin{tabular}{lcccc}
 & & CNN &  RES & ResBRN \\
 \cmidrule{3-5}
\multirow{2}{*}{CHiME-5} &  SAT &  85.4 & 85.1 & \textbf{85.0} \\
& Binaural & 55.6 & 55.8 & \textbf{54.4} \\\hline
\multirow{2}{*}{AMI-IHM}    & dev   & 29.4  & \textbf{28.1}  & 29.5 \\
                            & eval  & 30.1  & \textbf{29.1}  & 29.8 \\\hline
\multirow{2}{*}{AMI-MDM}    & dev   & 45.3  & 43.7  & \textbf{43.2} \\
                            & eval  & 49.0  & 47.6  & \textbf{46.9} \\\hline
\end{tabular}

\end{table}
\begin{table}
\caption{WER (\%) comparison for white noise data augmentation for binaural microphone.}
\label{t:wer4}
\centering

\begin{tabular}{lccccc}
 & & CNN &  RES & \begin{tabular}{@{}l@{}}RES \\ +Dropouts\end{tabular}  & ResBRN \\
 \cmidrule{3-6}
\multirow{2}{*}{CHiME-5} & SAT &  81.4 & 81.3 & 83.8 & \textbf{80.8} \\
& Binaural & \textbf{50.4} & 51.4 & 64.0 & 51.3 \\\hline
\end{tabular}
\end{table}

\begin{table}
\caption{WER (\%) comparison for the effectiveness of the multilevel LM.}
\label{t:wer3}
\centering
\begin{tabular}{lcccc}
 & & CNN &  RES & ResBRN \\
 \cmidrule{3-5}
\multirow{2}{*}{CHiME-5} &  SAT
                                 & 81.5 & 81.2 & \textbf{80.7} \\
                      & Binaural & \textbf{50.0} & 51.3 & 51.0 \\
\hline
\end{tabular}
\end{table}


\subsection{Residual Connections and Batch Renormalization}
Table~\ref{t:wer2} lists the WER for the CNN-based parallel encoder (CNN) added with residual connections (RES) and batch renormalization (ResBRN). 

For CHiME-5, the residual connections resulted in an additional absolute reduction of 0.3\% in the single-array track WER. 
After training the residual connections with batch renormalization, the joint model provided additional reductions of 0.1\% and 1.4\% on the single-array track and binaural tasks, respectively.

For AMI, the residual connections provided at least a reduction of 1.6\% of the WER. In addition, ResBRN reduced the WER by 0.5\% absolute for AMI-MDM. 

\subsection{Data Perturbation}
\label{sbsec:perturbation}
In addition to the abovementioned results, we report herein the WER for a model with a parallel encoder trained with augmented data on CHiME-5. The augmented data were obtained by adding simulated white noise to the binaural array. The SNR ratio was randomly selected to range from 7 to 20 dB. Table ~\ref{t:wer4} presents the resulting WER. ResBRN showed that the augmented data worked for the single-array track when noise was added to the binaural array. 
Adding dropouts in the residual connection led to a strong degradation because it affected both inputs of the parallel encoder, where the audio input from the single-array track was already degraded owing to the environmental setup.


\subsection{Multilevel LM}
Table~\ref{t:wer3} presents the WER for the multilevel LM used with a parallel encoder on CHiME-5. 
Using the parallel encoder resulted in the multilevel LM providing an additional 0.1\% improvement. In general, our final model with the proposed extensions performed better, providing absolute WER improvements of 14\% and 11\%, compared to the end-to-end and GMM baselines (Table~\ref{t:wersinglech}). The proposed extensions were able to overcome the results of the state-of-the-art lattice free MMI (LF-MMI) baseline without using any phonemic information or finite-state transducer decoding, and the results of the CMU proposal \cite{Dalmia}.  


\section{Conclusions}
\label{sec:conclusions}

We presented herein the extensions for a joint CTC--attention model based on residual learning, batch renormalization, and multilevel LM.
We applied a parallel encoder for multichannel input which accepts inputs with a different number of channels. To improve the processing of the audio features, we applied residual connections with batch renormalization. Then, we applied a multilevel LM which integrates the strength of a character-based LM and a word-based LM. 
Each extension improved the performance of the end-to-end models in everyday-environment ASR with respect to the single channel model and the end-to-end model proposed in \cite{Barker}, resulting in a WER absolute reduction of 8.5\% from the single channel end-to-end. However, it required the overall system to improve the WER with respect to the best baseline (LF-MMI TDNN) and it only obtained the reduction of 0.6\% absolute on the CHiME-5 corpus.

The proposed model employed 6 CNN layers and 4 RNN layers with 512 cells; however, due to the limitations of the GPU, very deep models were not possible to train without reducing the size of the mini-batch. The result obtained in training of deeper models and smaller mini-batch showed no improvement in the WER reduction. 
Furthermore, a training longer than 15 epochs did not show improvement on the accuracy or decreased the loss.  
The models showed improvements over the baseline even when no additional preprocessing, such as beamforming, was performed for the input.

\bibliographystyle{IEEEbib}
\bibliography{refs}

\end{document}